\documentclass[amsmath,nofootinbib,amsfonts,amssymb,aps,prd,12pt]{revtex4-1}
\usepackage{setspace}
\usepackage{subfigure}
\usepackage{graphicx}
\usepackage{color}
\usepackage{cancel}
\usepackage[colorlinks=true
,urlcolor=blue
,citecolor=blue
,linkcolor=blue
,pagecolor=blue
,linktocpage=true
,pdfproducer=medialab
]{hyperref}
\usepackage{comment}

\usepackage{longtable}
\usepackage{multirow}

\makeatletter

\newcommand{\Rmnum}[1]{\expandafter\@slowromancap\romannumeral #1@}
\makeatother
\begin{document}
\begin{flushright}
DESY 16-077\\
\end{flushright}


\begin{center}
\baselineskip 20pt {\Large\bf
Distinguishing Dirac/Majorana Sterile Neutrinos\\ at the LHC
}
\vspace{1cm}

{\large
Claudio O. Dib$^{a}$\footnote{E-mail: claudio.dib@usm.cl},
C. S. Kim$^{b}$\footnote{E-mail: cskim@yonsei.ac.kr},
Kechen Wang$^{cd}$\footnote{E-mail: kechen.wang@desy.de},
Jue Zhang$^{c}$\footnote{E-mail: zhangjue@ihep.ac.cn}
} \vspace{.5cm}

{\baselineskip 20pt \it $^a$
CCTVal and Department of Physics, \\
Universidad T\'ecnica Federico Santa Mar{\'{\i}}a, Valpara{\'{\i}}so, Chile \\
}
{\it $^b$
Department of Physics and IPAP, Yonsei University, Seoul 120-749, Korea \\
}
{\it $^c$
Institute of High Energy Physics, Chinese Academy of Sciences, Beijing, 100049, China \\
}
{\it $^d$
DESY, Notkestraße 85, D-22607 Hamburg, Germany \\
}

\vspace{.5cm}

\vspace{1.25cm} {\bf Abstract}
\end{center}
We study the purely leptonic decays of $W^\pm \to e^\pm  e^\pm  \mu^\mp \nu$  and $\mu^\pm  \mu^\pm  e^\mp \nu$ 
produced at the LHC, induced by sterile neutrinos with mass $m_N$ below $M_W$ in the intermediate state.
Since the final state neutrino escapes detection, one cannot tell whether this process violates  lepton number, what would indicate a Majorana character for the intermediate sterile neutrino. 
Our study shows that when the sterile neutrino mixings with electrons and muons are different enough, one can still discriminate between the Dirac and Majorana character of this intermediate neutrino by simply counting and comparing the above decay rates.
After performing collider simulations and statistical analysis, we find that at the $14~\text{TeV}$ LHC with an integrated luminosity of $3000~\text{fb}^{-1}$, for two benchmark scenarios $m_N$ = 20 GeV and 50 GeV, at least a $3\sigma$ level of exclusion on the Dirac case can be achieved for disparities as mild as e.g.  $|U_{Ne}|^2 < 0.7~ |U_{N\mu}|^2$ or $|U_{N\mu}|^2 < 0.7~ |U_{N e}|^2$, provided that $|U_{Ne}|^2$, $|U_{N\mu}|^2$ are both above $\sim 2\times 10^{-6}$.  


\thispagestyle{empty}

\newpage

\addtocounter{page}{-1}

\baselineskip 18pt



\section{Introduction}

Neutrinos are very special among the currently known elementary particles \cite{Agashe:2014kda}. They have tiny masses compared to the rest of the elementary fermions in all known processes. So far they are produced only at relativistic energies.
Their interactions are very weak, so that an overwhelming majority of them escape direct detection. Their flavor mixings turn out to be much larger that those of quarks. If there are right handed chiral components, they must be sterile under the known interactions. The masses of sterile components, if they exist, could be Dirac or Majorana and could be as large as GUT size. Large scale masses could explain the smallness of the known neutrino components by the so called
seesaw mechanisms (in several versions) \cite{see-saw}; there could be neutrino components that comprise the Dark Matter of the Universe \cite{DarkMatter}; and the CP violation in the lepton sector could provide explanation for the baryon asymmetry of the universe \cite{BA-Lep}.

In this work, we address a simple way to discriminate the Dirac vs. Majorana character of sterile neutrinos, provided they exist with masses near and below the $W$ boson mass. Currently, the main experiments that are sensitive to the Majorana character of neutrinos are those that search for neutrinoless double beta decays ($0\nu\beta\beta$) \cite{0nuBB}. These experiments are sensitive to Majorana neutrinos, in principle of any masses, including the known light neutrinos. However, the extraction of parameters from $0\nu\beta\beta$ experiments will not be an easy task, due to at least two reasons: firstly, there are large theoretical uncertainties in the estimation of the nuclear matrix elements involved; secondly, there could be cancellations of interfering amplitudes which will not be possible to disentangle without further inputs from other experiments. Therefore, it can be important to have additional ways, such as BaBar \cite{BABAR:2012aa, Lees:2013gdj},  BELLE \cite{Seon:2011ni, Petric:2011zz} and LHCb \cite{Adeva:2013csa, Khanji:2014dca} studies on rare lepton flavor and lepton number violating  decays of heavy mesons \cite{Cvetic:2010rw, Helo:2010cw}, to discriminate between the Dirac vs. Majorana character of neutrinos. 

One such additional way is the search for equal sign dileptons in high energy colliders, which are suitable for large neutrino masses. Indeed, for neutrino masses above $\sim 100$ GeV the LHC experiments can use the mode $pp \to \ell^\pm \ell^\pm j j$ \cite{Aad:2015xaa, Khachatryan:2015gha,Helo:2013esa, Keung:1983uu}. On the other hand, for neutrino masses below $M_W$ the produced jets in the final state $\ell^\pm \ell^\pm j j$ may not pass the cuts required to reduce backgrounds, so that purely leptonic modes such as $pp   \to \ell^\pm \ell^\pm \ell^{\prime\mp}\nu$ can be more favorable \cite{Izaguirre:2015pga}.
In our previous work \cite{Dib:2015oka} we studied the signal $W^\pm \to e^\pm e^\pm \mu^\mp \nu$, which will appear resonantly enhanced provided there exist neutrinos with masses below $M_W$, through the subprocess $W^\pm \to e^\pm N$ followed by
$N\to e^\pm\mu^\mp\nu$. The choice of having no opposite-sign same-flavor (no-OSSF) lepton pairs in the final state helps eliminate a serious standard model radiative background  $\gamma^*/Z \to \ell^+\ell^-$ \cite{p2mu}.
Now, a heavy neutrino $N$, if it is Majorana,  will induce the lepton number conserving (LNC) $W^+ \to e^+ e^+ \mu^- \nu_e$  as well as the lepton number violating (LNV) $W^+ \to e^+ e^+ \mu^- \bar\nu_\mu$ process, while if it is of Dirac type, it will induce only the LNC process (for simplicity of notation we considered the $W^+$ decays, but the same is valid for the $W^-$).
Consequently, one could in principle try to observe a difference in these processes that could test whether the neutrino $N$ is Majorana or Dirac.
However, since the final neutrino escapes detection, the observed final state is just $e^\pm e^\pm \mu^\mp$ or $\mu^\pm \mu^\pm e^\mp$ { plus missing energy. Hence, apparently, it is not a simple task} to distinguish between the LNC and the LNV processes.
Therefore, although this tri-lepton channel is appropriate for searching sterile neutrinos with masses below $M_W^{}$,  the determination of the Dirac or Majorana character of the sterile neutrino is  quite challenging.

In our previous work \cite{Dib:2015oka} we found that one could still discriminate the Dirac vs. Majorana nature of the heavy neutrino by studying the energy distribution of the lepton in the final state with charge opposite to the decaying $W$ boson
(e.g.  $\mu^-$ in the process $W^+ \to e^+ e^+ \mu^- \nu$). 
In this work, we present a simpler method to distinguish between Majorana and Dirac $N$ by examining the integrated decay rates, not the spectra, for all the channels $e^\pm e^\pm \mu^\mp$ and $\mu^\pm \mu^\pm e ^\mp$, because the discrimination through the spectra could be a highly challenging task for experiments. This method, on the other hand, is useful only in the case that  the mixing parameters $U_{Ne}$ and $U_{N\mu}$ are considerably different from each other.

This method works due to the fact that, for Dirac sterile neutrinos, no matter whether the mixing angles with electrons and muons are equal or not, the decays into the channels $e^\pm e^\pm \mu^\mp$ and $\mu^\pm \mu^\pm e ^\mp$ are all LNC processes and are all the same, while for Majorana sterile neutrinos, with different mixing parameters there is always one LNV process which dominates over the LNC processes. Consequently, if the true nature of sterile neutrinos is of Majorana type and their mixing angles with electrons and muons are different enough, the observed number of events in the $e^\pm e^\pm  \mu^\mp$ and $\mu^\pm \mu^\pm e^\mp$ channels will be different, unlike the case for Dirac sterile neutrinos. Details of the method are described in the next section.

It should be noted that such a finding is only valid at the theoretical level. In an actual collider search, because of the detector effects, the observed numbers of events in the $e^\pm_{} e^\pm_{} \mu^\mp_{}$ and $\mu^\pm_{} \mu^\pm_{} e^\mp_{}$ channels could be different even for the Dirac sterile neutrino case. 
Therefore, it is worthwhile to perform a careful collider simulation and carry out a detailed statistical study for this scenario, which we present in detail in Section \ref{sec:simulation}.
Now, in the less fortunate scenario where $N$ has equal --or nearly equal-- mixing parameters with electrons and muons, the above method no longer applies, as both Dirac or Majorana neutrino will give the same number of events in the above two channels, in which case one can only resort to the spectral distributions of final state leptons, as shown in the Ref. \cite{Dib:2015oka}. A further investigation along this line has also been pursued by us recently, and the obtained results will be presented elsewhere \cite{DKWZ}.

The rest of the paper is organized as follows. In Section \ref{sec:rate}, we review the expressions for the decay branching ratios of $W$ to tri-leptons via the sterile neutrino $N$. The detailed collider simulation for the scenario with disparate mixing is carried out in Section \ref{sec:simulation}, followed by a statistical analysis on excluding the Dirac sterile neutrino case in Section \ref{sec:stat}. We conclude in Section \ref{sec:summary}. 

\section{Leptonic Decay of $W$ via Sterile Neutrino $N$}
\label{sec:rate}

\begin{figure}[h]
\centering
\begin{minipage}[b]{.35\linewidth}
 \includegraphics[width=\linewidth]{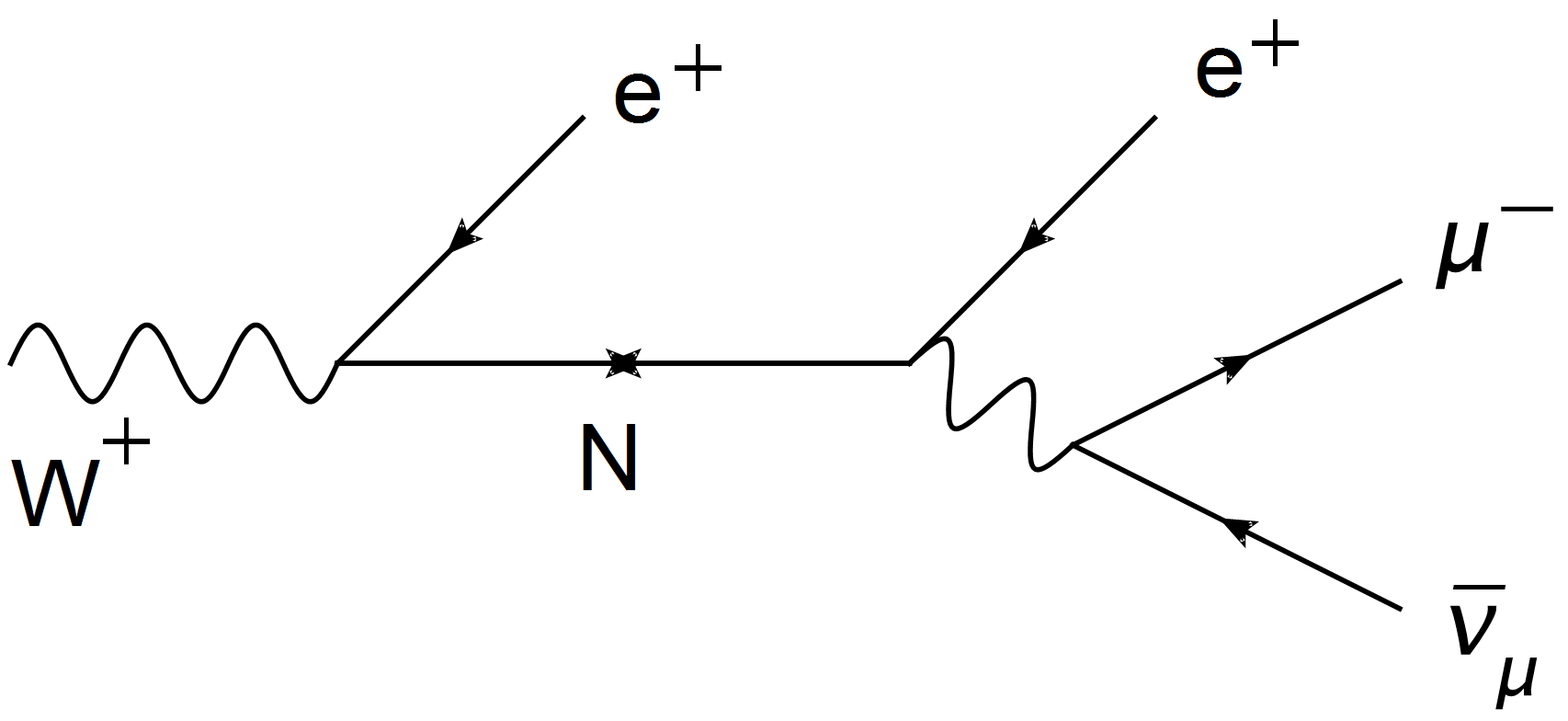}
\end{minipage}
\hspace{18pt}
\begin{minipage}[b]{.35\linewidth}
\vspace{0pt}
 \includegraphics[width=\linewidth]{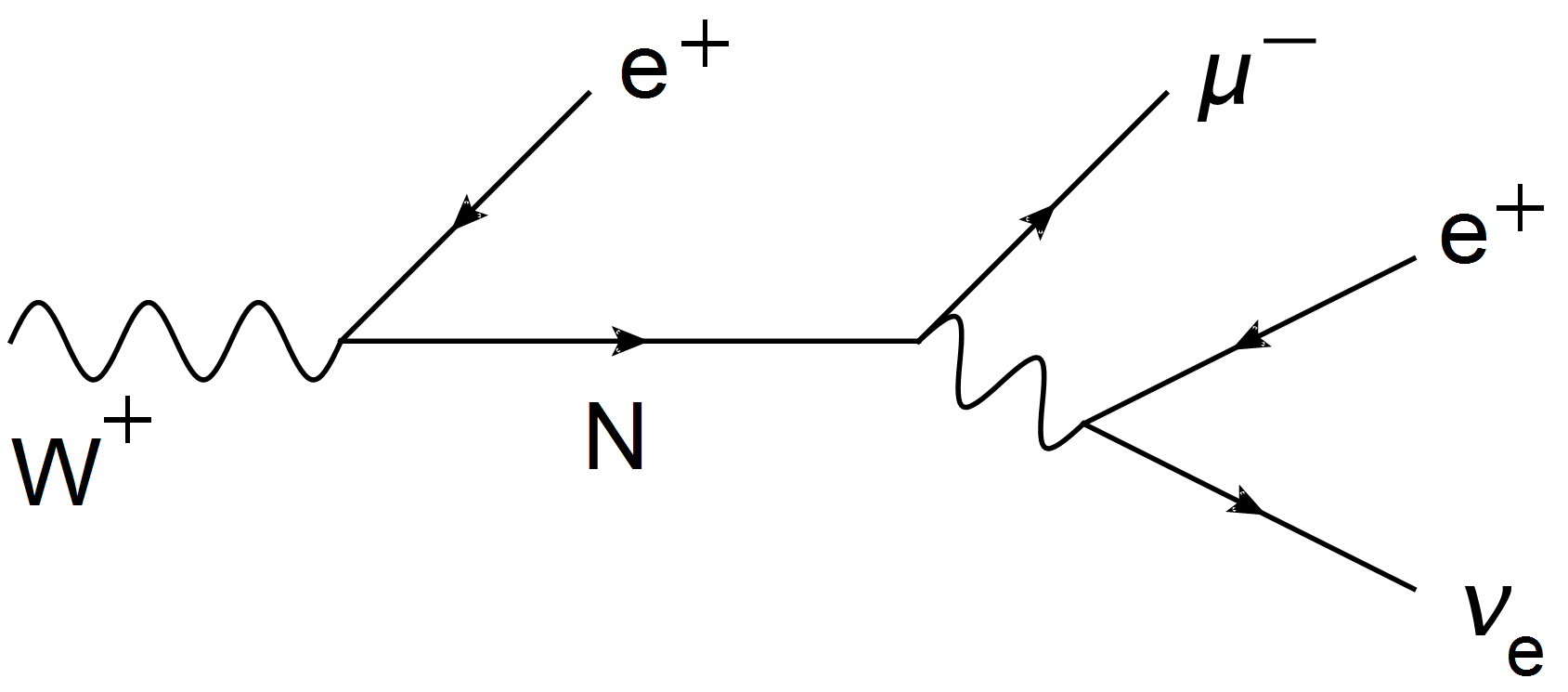}
\end{minipage}
\vspace{-0.2cm} \caption{ The lepton number violating (LNV) process $W^+\to e^+ e^+ \mu^-\bar\nu_\mu$,  mediated by a heavy sterile neutrino of Majorana type (left) and the lepton number conserving (LNC) process $W^+\to e^+ e^+ \mu^- \nu_e$, mediated by a heavy sterile neutrino of Majorana or Dirac type (right).} \label{fig1}
\end{figure}

As mentioned in the Introduction section, we will study decays into final states without dileptons of the same flavor and opposite sign in order to avoid large SM backgrounds from radiative pair production, namely we consider $W^+ \to e^+ e^+ \mu^{-} 
{\nu}$ and $W^+ \to \mu^+ \mu^+ e^{-} 
{\nu}$ (or their charge conjugates) but neither
$W^+ \to e^+ e^{-} \mu^+ 
{\nu} $ nor $W^+ \to \mu^+ \mu^{-} e^+ 
{\nu}$. For illustration, in Fig.~\ref{fig1} we show the corresponding Feymann diagrams for the LNC and LNV processes in the $e^+_{} e^+_{} \mu^-_{}$ channel.

For the LNV processes $W^+ \to e^+ e^+ \mu^{-} \bar\nu_\mu$ and $W^+ \to \mu^+ \mu^+ e^{-} \bar\nu_e$, the rates are: 
\begin{eqnarray}
\Gamma (W^+ \to e^+ e^+\mu^-\bar\nu_\mu) &=&
 |U_{N e}|^4 \times \hat\Gamma \ \  \propto \ \  |U_{N e}|^4, \nonumber \\
\Gamma (W^+ \to \mu^+ \mu^+ e^{-} \bar\nu_e) &=&
 |U_{N \mu}|^4 \times \hat\Gamma \ \ \propto\ \  |U_{N \mu}|^4.
\label{LNV_rate}
\end{eqnarray}
where the factor $\hat\Gamma$ is
\begin{equation}
\hat\Gamma = \frac{G_F^3 M_W^3}{12\times 96 \sqrt{2} \ \pi^4}  \frac{ m_N^{5}}{\Gamma_N}
\left( 1-\frac{m_N^2}{M_W^2} \right)^2    \left(1 + \frac{m_N^2}{2 M_W^2} \right).
\end{equation}
$G_F^{}$ is the Fermi's constant and $M_W^{}$ is the $W$-boson mass. Here we have assumed that only a single sterile neutrino $N$ participates in the decay of $W$, with its total decay width denoted as $\Gamma_N^{}$, its mass as $m_N^{}$, and its mixing with a charged lepton $\ell$ denoted as $U_{N \ell}$. Notice that, although the processes in Eq.~\ref{LNV_rate} appear to be quartic in the mixings $U_{N\ell}$, they are effectively quadratic when $N$ goes on its mass shell, because $\Gamma_N$ is quadratic in $U_{N\ell}$.

On the other hand, the corresponding rates for the LNC processes $W^+ \to e^+ e^+ \mu^{-} \nu_e$ and $W^+ \to \mu^+ \mu^+ e^{-} \nu_\mu$ are given by the similar expressions:
\begin{eqnarray}
\Gamma (W^+\to e^+ e^+\mu^-\nu_e) &=&
 |U_{N e} U_{N\mu}|^2 \times \hat\Gamma \ \  \propto\ \  |U_{N e} U_{N\mu}|^2 \nonumber, \\
\Gamma (W^+\to \mu^+ \mu^+ e^{-} \nu_\mu) &=&
 |U_{N e} U_{N\mu}|^2\times \hat\Gamma  \ \ \propto\ \  |U_{N e} U_{N\mu}|^2.
\label{LNC_Rate}
\end{eqnarray}

The expressions differ only in the lepton mixing factors: $|U_{N e}|^4$ for the LNV process
$W^+ \to e^+ e^+ \mu^{-} \bar\nu_\mu$, $|U_{N \mu}|^4$ for  the LNV process $W^+ \to \mu^+ \mu^+ e^{-} \bar\nu_e$,
and $|U_{N e} U_{N \mu}|^2$ for the two LNC processes  $W^+ \to e^+ e^+ \mu^{-} \nu_e$ and $W^+ \to \mu^+ \mu^+ e^{-} \nu_\mu$. So, clearly both LNC processes have always equal rates, while the LNV processes are equal only if $|U_{Ne}|^2 =|U_{N\mu}|^2$. With the fact that a Dirac sterile neutrino will produce only the LNC processes while a Majorana neutrino will produce both the LNC and LNV processes, one can compare the production of $e^+ e^+ \mu^-$ and $\mu^+ \mu^+ e^- $ (or their charge conjugates) induced by a Dirac or a Majorana sterile neutrino.

\begin{table}[h]
\centering
\begin{tabular}{c | c | c}
\hline
\hline
 & Dirac & Majorana \\
\hline
$e^+ e^+ \mu^-$ & $1$ & $1+r$ \\
$\mu^+ \mu^+ e^-$ & $1$ & $1+1/r$ \\
\hline
\hline
\end{tabular}
\caption{Relative factors in the branching ratios of $W^+ \rightarrow e^+ e^+ \mu^- {\nu}$ and $W^+ \rightarrow \mu^+ \mu^+ e^- {\nu}$ for both Dirac and Majorana sterile neutrino scenarios, where ${\nu}$ represents a standard neutrino or anti-neutrino. The same applies for the respective charge conjugate modes. Here $r$ is defined as $r \equiv  |U_{Ne}^{}|^2/|U_{N\mu}^{}|^2$.}
\label{tb:comparison}
\end{table}

In Table \ref{tb:comparison}, we present such a comparison, where the rate of $W^+ \rightarrow e^+ e^+ \mu^- \nu$ in the Dirac case is chosen as the reference value, by which the rates of other cases are normalized. It is now apparent that in the case of a Dirac sterile neutrino the production rates of $e^+_{} e^+_{} \mu^-_{}$ and $\mu^+_{} \mu^+_{} e^-_{}$ should be equal, while for the Majorana case they will differ, depending on the \emph{disparity factor}, $r$,  defined as:
\begin{equation}
 r \equiv \frac{ |U_{Ne}^{}|^2}{|U_{N\mu}^{}|^2}. 
 \label{disparity}
 \end{equation}
 For $r> 1$, the number of  $e^+_{} e^+_{} \mu^-_{}$ events should be 
larger than the $\mu^+_{} \mu^+_{} e^-_{}$ events, and viceversa, if $r < 1$ the $\mu^+_{} \mu^+_{} e^-_{}$ events will be more abundant than $e^+_{} e^+_{} \mu^-_{}$. Similar comparison also exists in the corresponding charge-conjugated processes.

The essence of the above feature can be attributed to the requirement of having no lepton pairs with  opposite sign and same flavor (no-OSSF) in the final state. With such requirement, the diagrams of the LNV and LNC processes have different topological structures. As shown in Fig.~\ref{fig1}, in the LNC process the fermion line containing the sterile neutrino $N$ must be attached to final leptons of opposite charge, and consequently these must have different flavors,
hence the mixing factor $|U_{Ne} U_{N\mu}|$.
On the other hand, in the LNV process, the sterile neutrino line is attached to two leptons of same sign and same flavor, hence the factor $|U_{N e}|^4$. Actually, since this difference is valid irrespectively of the mass of $N$, one may generalize our current study to the case where $m_N^{} > M_W^{}$, although for larger masses the dilepton-dijet processes $\ell\ell j j$ are favored as they tend to give larger rates even after cuts to reduce the hadronic background.

Consequently, in the following section we will restrict our study to the case where $m_N^{} < M_W^{}$, 
for different values of the $r$ parameter (see Table \ref{tb:comparison}).
We will perform the statistical analysis with collider simulations for both background and signals, and determine the statistical level at which one can distinguish a Majorana vs. a Dirac sterile neutrino case at the LHC.

\section{Collider Simulation}
\label{sec:simulation}

In our simulation, we first build a Universal FeynRules Output \cite{Degrande:2011ua} model file using $\texttt{FeynRules}$ \cite{Christensen:2008py} which extends the SM model with additional sterile neutrino interactions. Both signal and background events are generated within the framework of $\texttt{MadGraph 5}$ \cite{Alwall:2014hca}, where the parton showering and detector simulation are carried out by $\texttt{PYTHIA 6}$ \cite{Sjostrand:2006za} and $\texttt{DELPHES 3}$ \cite{deFavereau:2013fsa}, respectively. At the parton level, we include up to two extra partons for both signal and background processes, and perform the jet matching using the MLM-based shower-$k_\bot^{}$ scheme \cite{Alwall:2008qv}. Lastly, to maintain consistency across the processes we are considering, we present the results using the cross sections from the $\texttt{MadGraph 5}$ output.

Although in this trilepton search we demand no OSSF lepton pairs in the final state, there still exists non-negligible background from various processes. We divide them into two categories. In the first category, we have the pair production of of $WZ$ with $W$ decaying leptonically and $ Z \rightarrow \tau^+_{} \tau^-_{}$. The subsequent decay of the $\tau$'s can lead to trilepton events with no OSSF lepton pairs. We estimate this background process via the Monte Carlo simulation.

The second category of background consists in ``fake" leptons which mainly originate from heavy-flavor meson decays. Although in general leptons from such a heavy-flavor meson decay are not well isolated, there are still rare occasions when they can pass the lepton isolation criteria \cite{Sullivan:2006hb,Sullivan:2008ki,Sullivan:2010jk}.  Dominant background processes of this kind are $\gamma^*_{}/Z$+jets and $t\bar{t}$, where an event with no OSSF lepton pairs
arises from $\gamma^*_{}/Z \rightarrow \tau^+_{} \tau^-_{}$ or the prompt decay of $t$ and $\bar{t}$, and a third lepton is faked from jets containing heavy-flavor mesons. Because these processes have large cross sections and small fake probabilities, it is very challenging to obtain enough statistics for background study in the pure MC simulation. Moreover, simulating such processes requires a detailed modelling of the jet fragmentations, and current level of MC simulation may not be accurate enough. For these reasons, data-driven methods are used by the ATLAS and CMS collaborations to estimate the fake lepton contributions \cite{Khachatryan:2015gha,Chatrchyan:2014aea,ATLAS:2014kca}.

In this work, we adopt a phenomenological Fake Lepton (FL) simulation method originally introduced in Ref.~\cite{Curtin:2013zua} and later also implemented in
Ref.~\cite{Izaguirre:2015pga}. In this FL simulation, one employs the fact that FL's originate from jets, and therefore they inherit parts of the kinematics of the original jets. Two modelling functions are introduced: one is called ``mistag efficiency'', $\epsilon_{j \rightarrow \ell}^{}$, which represents the probability of a particular jet faked into a lepton, and the other is called ``transfer function", $\mathcal{T}_{j\rightarrow \ell}^{}$, which is a probability distribution that determines how much the jet momentum is transferred into the faked lepton. These two functions consist of a few modelling parameters, which can be pinned down by validating simulated results against actual experimental ones. We revisit the validation performed in Ref.~\cite{Izaguirre:2015pga}, and find that the modelling parameters they obtained can be consistent with the experimental results. Thus, the same set of parameters are used here, and we also assume the same fake efficiency for electrons and muons. Details of this FL simulation method and the validation can be found in Refs.~\cite{Izaguirre:2015pga,Curtin:2013zua}. Our validation results are given in Appendix \ref{sec:FL}.

We now discuss the various kinematic cuts applied in this simulation study.
At first, we impose the basic cuts for leptons and jets: for leptons, $p_T^\ell \geq 10~\text{GeV}$ and rapidity $|\eta_\ell^{}| \leq 2.5$; and for jets, $p_T^j \geq 20~\text{GeV}$ and rapidity $|\eta_j^{}| \leq 5.0$.

\begin{figure}[h]
\centering
\includegraphics[scale=0.85]{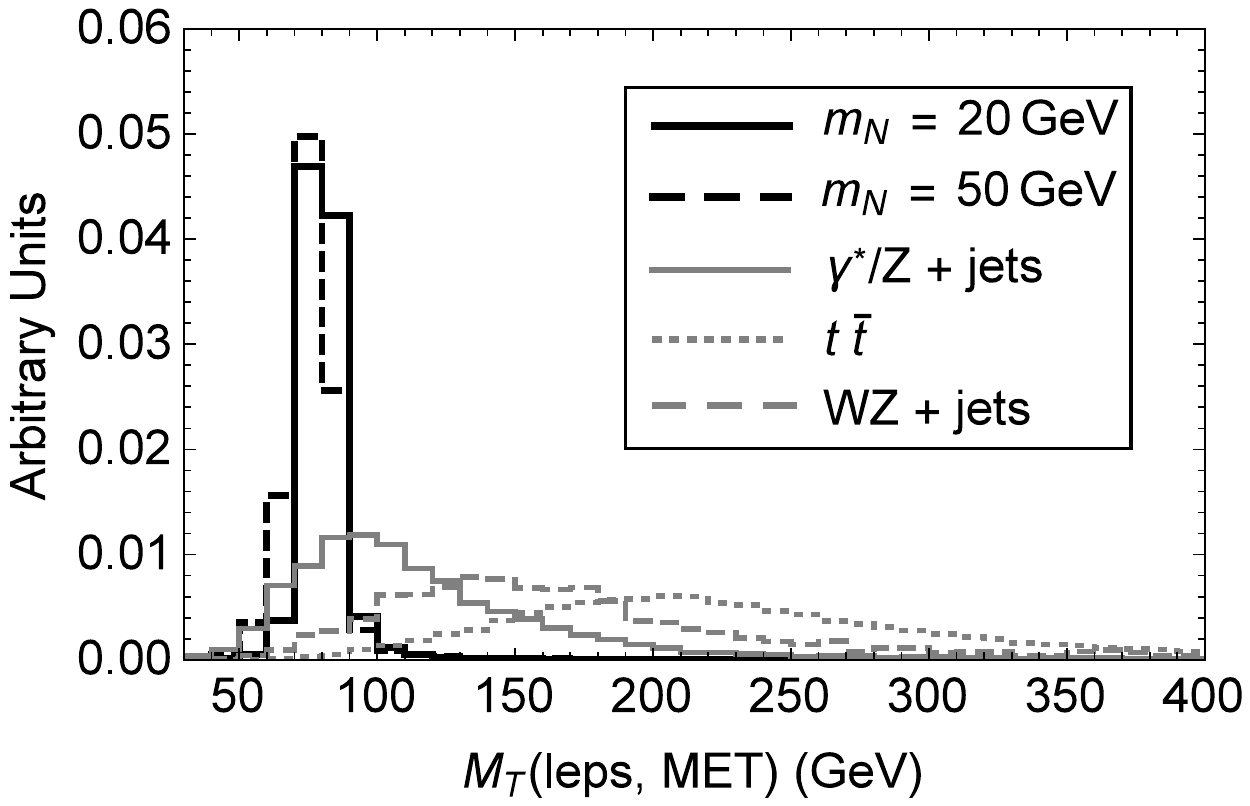}
\caption{Distributions of $M_T^{}(\text{leps}, \cancel{E}_T^{})$ for both signal and background processes after basic cuts. For signal processes, we choose the LNC case in the $e^+_{} e^+_{} \mu^-_{}$ channel as an example. 
}
\label{fg:MtLepsMET}
\end{figure}

We then require the transverse mass $M_T^{}(\text{leps}, \cancel{E}_T^{}) < 90~\mathrm{GeV}$, with $M_T^{}(\text{leps}, \cancel{E}_T^{})$ defined as
\begin{eqnarray}
M_T^{2}(\text{leps}, \cancel{E}_T^{}) = m_{\text{leps}}^{2} + 2 \left( \sqrt{\left(m_{\text{leps}}^{2} +\vec{p}_{\text{leps},T}^{2} \right) ~ \vec{\cancel{p}}_T^{2} } - \vec{\cancel{p}}_T^{} \cdot \vec{p}_{\text{leps},T}^{} \right),
\end{eqnarray}
where $m_{\text{leps}}^{}$ and $ \vec{p}_{\text{leps},T}^{}$ are the invariant mass and transverse momentum of all three visible leptons in the final state, and $\vec{\cancel{p}}_T^{}$ is the missing transverse momentum. For the signal processes, this $M_T^{}$ is bounded by the W-boson mass, because all the charged leptons and neutrinos in the final state come from the decay of a W-boson. Considering the detector resolution and the W-boson width, we find that  $M_T^{}(\text{leps}, \cancel{E}_T^{}) < 90~\mathrm{GeV}$ is an good discriminating condition between the signal and background, especially in the elimination of large background from $t\bar{t}$. For illustration, in Fig.~\ref{fg:MtLepsMET}, we show the distributions of $M_T^{}(\text{leps}, \cancel{E}_T^{})$ after basic cuts for both background and signal processes in the $e^+_{} e^+_{} \mu^-_{}$ channel.

To further suppress background, we impose the cuts $\cancel{E}_T^{} < 40~\mathrm{GeV}$, zero b-jets, and  $H_T^{} < 50~\mathrm{GeV}$, where $H_T^{}$ is the scalar sum of all jets $p_T^{j}$.
\begin{figure}[h]
\centering
\includegraphics[scale=0.65]{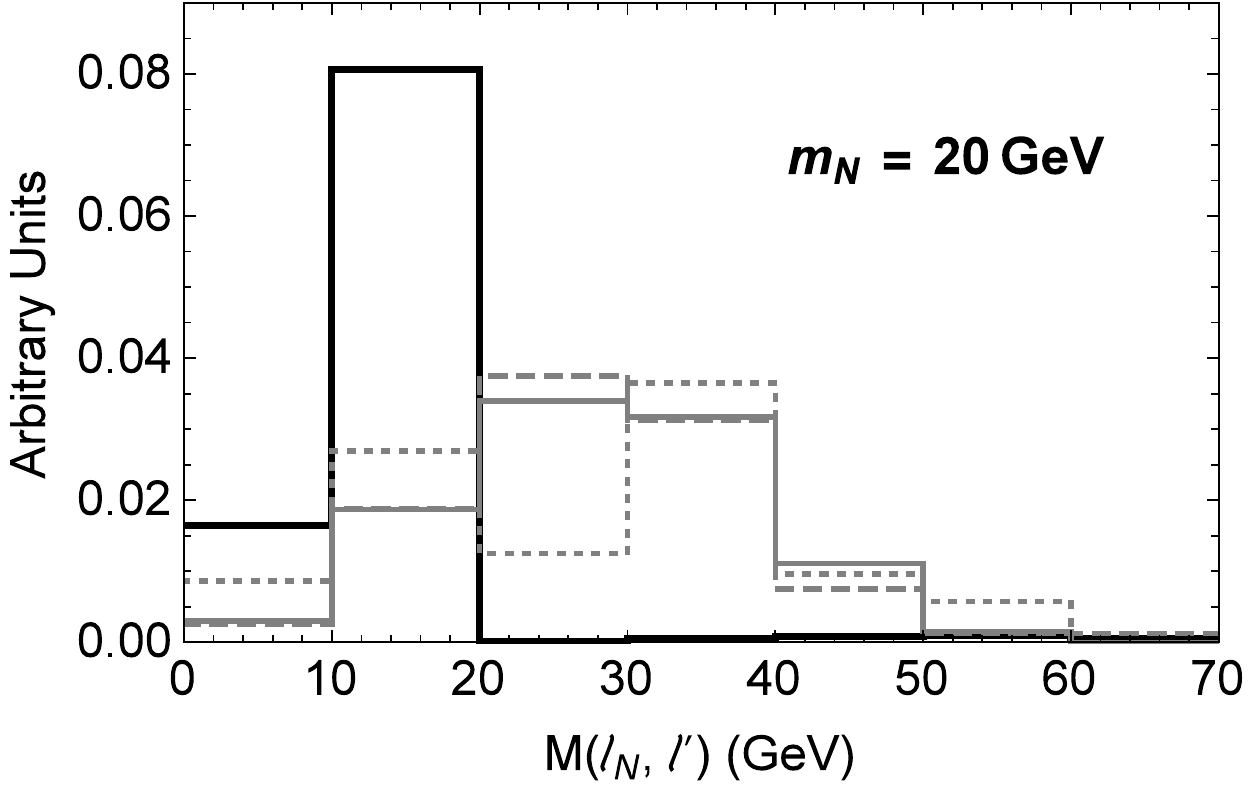}
\includegraphics[scale=0.65]{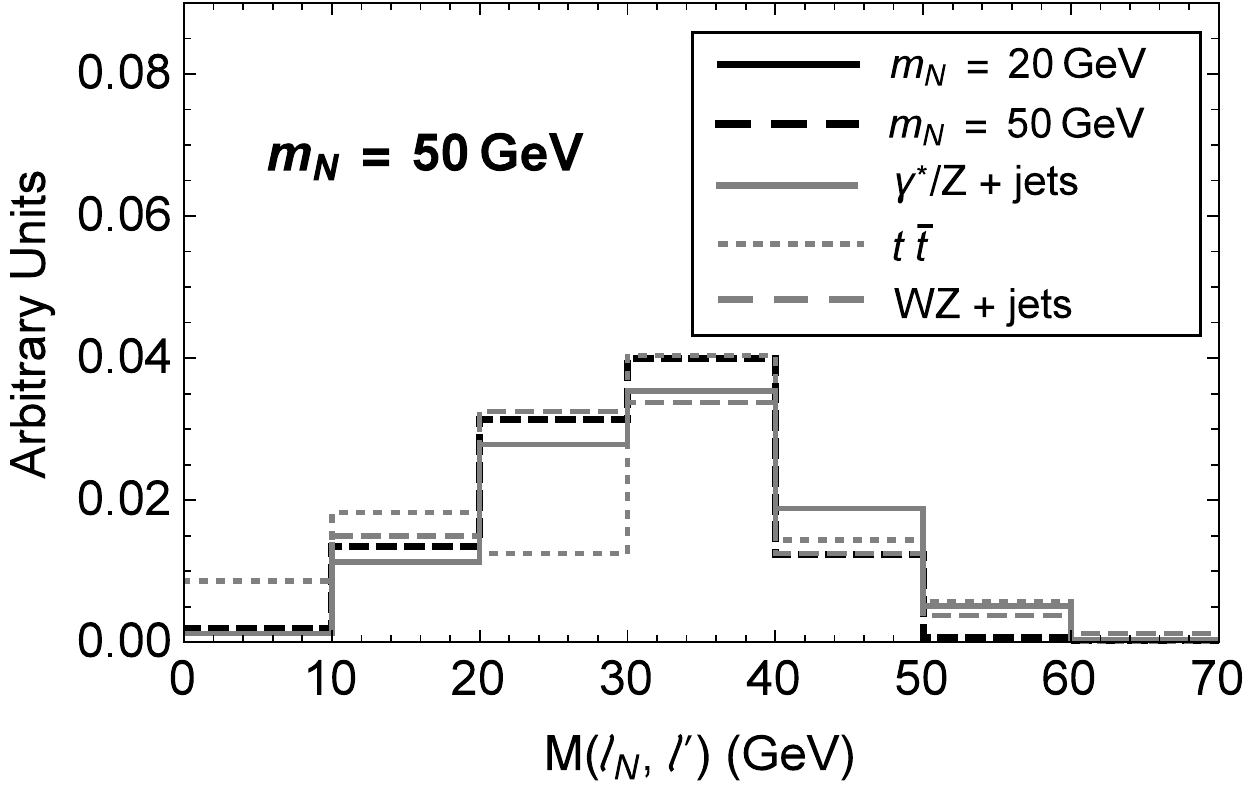}
\caption{Distributions of $M(\ell_N^{},\ell^\prime)$ for both signal and background processes after applying all cuts in Table \ref{tb:bkgd} except for the last one. For signal processes, we choose the LNC case in the $e^+_{} e^+_{} \mu^-_{}$ channel as an example.}
\label{fg:MdLep2sLep}
\end{figure}

The above cuts are proved to be very effective in reducing the background from $WZ$ and $t \bar t$.
 However, the background from $\gamma^*/Z$+jets is still large (see Table \ref{tb:bkgd}). As was also observed in Ref.~\cite{Izaguirre:2015pga}, for the $m_N^{} = 20~{\text{GeV}}$ case one is able to impose a further cut using the fact that the invariant mass of the two leptons originating from the $N$ decay should be bounded from above by  the mass of $N$. Such a cut requires a correct identification of the lepton pair, as there will be two final state leptons which are indistinguishable by their signs and flavors. In Appendix \ref{sec:SL_order}  we provide a method for identifying the correct lepton pair. The three leptons in the final state are labeled as $\ell_
W^\pm \ell_N^\pm \ell^{\prime \mp}$, with $\ell_W^{}$ and $\ell_N^{}$ denoting the leptons from the prompt decays of $W$ and $N$, respectively. Note that this additional cut on the invariant mass of $M(\ell_N^{}, \ell_{}^{\prime})$ is effective in the $m_N^{} = 20~{\text{GeV}}$ case while not as much in the $m_N^{} = 50~{\text{GeV}}$ case. This can be seen from Fig.~\ref{fg:MdLep2sLep}, where the distributions of $M(\ell_N^{},\ell^\prime)$ for both signal and background processes are shown. As one can see, for the $m_N^{} = 20~{\text{GeV}}$ case one can further reduce the background by requiring $M(\ell_N^{}, \ell_{}^{\prime}) < 20~\text{GeV}$, however, a similar cut cannot be applied to the $m_N^{} = 50~{\text{GeV}}$ case, as the signal process exhibits almost the same distribution as the background ones.

\begin{table}[h]
\centering
\begin{tabular}{c | c | c | c | c}
\hline
\hline
\multirow{2}{*}{Cuts} & \multicolumn{2}{c|}{$WZ$} & $\gamma^* / Z$+jets &  $t\bar{t}$ \\
\cline{2-5}
& $\ell^+_{} \ell^+_{} \ell^{\prime -}_{}$ & $\ell^-_{} \ell^-_{} \ell^{\prime +}_{}$ & $\ell^\pm_{} \ell^\pm_{} \ell^{\prime \mp}_{}$ & $\ell^\pm_{} \ell^\pm_{} \ell^{\prime \mp}_{}$ \\
\hline
Basic cuts & 779 & 550 & 1055 & 17147 \\
\hline
$M_T^{}(\text{leps}, \cancel{E}_T^{}) < 90~\mathrm{GeV}$ & 52 & 34 & 374 & 160 \\
\hline
$\cancel{E}_T^{} < 40~\mathrm{GeV}$ & 46 & 28 & 356 & 113 \\
\hline
$N(\text{b-jets}) = 0, H_T^{} < 50~\mathrm{GeV}$ & 39 & 23 & 323 & 15 \\
\hline
$M(\ell_N^{}, \ell^\prime) <20~\mathrm{GeV}$ & 7.4 & 4.4 & 62 & 2.7 \\
\hline
\hline
\end{tabular}
\caption{Cut flow for background processes. Numbers of events correspond to an integrated luminosity of $3000~\mathrm{fb}^{-1}$ at the $14~\mathrm{TeV}$ LHC; $\ell$ includes both $e$ and $\mu$. 
}
\label{tb:bkgd}
\end{table}

In Tables \ref{tb:bkgd}, \ref{tb:mN20} and \ref{tb:mN50}  we show the cut flow tables for all the background processes, and two benchmark signal scenarios with $m_N^{} = 20~\text{GeV}$ and $50~\text{GeV}$, respectively. The numbers of events are calculated for an integrated luminosity of $3000~\text{fb}^{-1}_{}$ at the $14~\text{TeV}$ LHC. For $\gamma^*/Z$+jets and $t\bar{t}$, although the FL simulation method is adopted, it is still difficult to obtain enough statistics to resolve the small difference between the modes $e^\pm e^\pm \mu^\mp$ and $\mu^\pm \mu^\pm  e^\mp$. Hence we choose to combine them.  The symbol $\ell$ in Table \ref{tb:bkgd} includes both $e$ and $\mu$.  The same treatment is also made for $WZ$, where $\ell^+_{} \ell^+_{} \ell^{\prime -}_{}$ and $\ell^-_{} \ell^-_{} \ell^{\prime +}_{}$ are separated due to the different production rates of $W^+_{}Z$ and $W^-_{}Z$. For signal processes we distinguish not only different tri-lepton modes but also the LNC and LNV processes of interest here. 
For illustration purposes, we take $|U_{N e}|^2=|U_{N \mu}|^2 = 1\times 10^{-6}$ and $U_{\tau N}^{} = 0$. For other values of the mixings, the number of events can be scaled accordingly. It is worth noting that from Table \ref{tb:mN20} and \ref{tb:mN50}, the number of events for LNV process is always larger compared with LNC process after applying basic cuts. This is due to the fact that kinematical distrubtions of final state leptons are different between LNV and LNC processes. The leptons in the LNV process are more efficient to pass the basic selection cuts.

\begin{table}[h]
\centering
\begin{tabular}{c | c | c | c | c | c | c | c | c}
\hline
\hline
\multirow{2}{*}{Cuts} & \multicolumn{2}{c |}{$e^+ e^+ \mu^-$} & \multicolumn{2}{c |}{$\mu^+ \mu^+ e^-$} &  \multicolumn{2}{c |}{$e^- e^- \mu^+$} & \multicolumn{2}{c }{$\mu^- \mu^- e^+$}\\
\cline{2-9}
& LNC & LNV & LNC & LNV & LNC & LNV & LNC & LNV \\
\hline
Basic cuts & 13.6	 & 19.5 & 15.0 & 22.0 & 12.1 & 18.2 & 13.3 & 19.5 \\
\hline
$M_t(\text{leps, MET}) < 90~\mathrm{GeV}$ & 12.7 & 18.3 & 13.9 & 20.3 & 11.3 & 17.0 & 12.3 & 18.3 \\
\hline
$\text{MET} < 40~\mathrm{GeV}$ & 12.5 & 18.3 & 13.8 & 20.3 & 11.2 & 17.0 & 12.3 & 18.3 \\
\hline
$N(\text{b-jets}) = 0, H_t < 50~\mathrm{GeV}$ & 11.1 & 16.6 & 12.2 & 18.5 &  10.0 & 15.6 & 11.0 & 16.6\\
\hline
$M(\ell_N^{}, \ell^\prime) <20~\mathrm{GeV}$ & 10.8 & 16.3 & 11.8 & 17.8 & 9.8 & 15.1 & 10.7 & 16.1 \\
\hline
\hline
\end{tabular}
\caption{Cut flow for signal processes with $m_N^{} = 20~\mathrm{GeV}$, $|U_{N e}|^2=|U_{N \mu}|^2 = 1\times 10^{-6}$ and $U_{\tau N}^{} = 0$. Numbers of events correspond to an integrated luminosity of $3000~\mathrm{fb}^{-1}$ at the $14~\mathrm{TeV}$ LHC.
}
\label{tb:mN20}
\end{table}

\begin{table}[h!]
\centering
\begin{tabular}{c | c | c | c | c | c | c | c | c}
\hline
\hline
\multirow{2}{*}{Cuts} & \multicolumn{2}{c |}{$e^+ e^+ \mu^-$} & \multicolumn{2}{c |}{$\mu^+ \mu^+ e^-$} &  \multicolumn{2}{c |}{$e^- e^- \mu^+$} & \multicolumn{2}{c }{$\mu^- \mu^- e^+$}\\
\cline{2-9}
& LNC & LNV & LNC & LNV & LNC & LNV & LNC & LNV \\
\hline
Basic cuts & 27.7	 & 30.7 & 30.7 & 33.3 & 23.7 & 26.6 & 26.3 &	29.8 \\
\hline
$M_t(\text{leps, MET}) < 90~\mathrm{GeV}$ & 26.4 & 29.0 & 29.2 & 31.7 & 22.5 & 25.1 & 25.0 & 28.1 \\
\hline
$\text{MET} < 40~\mathrm{GeV}$ & 26.1 & 28.7 & 28.9 & 31.4 & 22.3 & 25.1 & 24.8 & 28.1 \\
\hline
$N(\text{b-jets}) = 0, H_t < 50~\mathrm{GeV}$ & 23.7 & 26.0 & 26.2 & 28.4 & 20.1 & 22.8 & 22.4 & 25.5\\
\hline
\hline
\end{tabular}
\caption{Cut flow for signal processes with $m_N^{} = 50~\mathrm{GeV}$, $|U_{N e}|^2=|U_{N \mu}|^2 = 1\times 10^{-6}$ and $U_{\tau N}^{} = 0$. Numbers of events correspond to an integrated luminosity of $3000~\mathrm{fb}^{-1}$ at the $14~\mathrm{TeV}$ LHC.}
\label{tb:mN50}
\end{table}

\section{Statistical Analysis}
\label{sec:stat}

We now turn to a detailed statistical analysis of our scenario of interest, where the sterile neutrino mixings $U_{Ne}^{}$ and $U_{N\mu}^{}$ are different. In the analysis, $N$ is assumed to be of Majorana type, and the question is whether it can be distinguished from a Dirac neutrino. As mentioned in Section \ref{sec:rate}, in this scenario one is able to distinguish Dirac from Majorana sterile neutrinos by comparing the numbers of $e^\pm e^\pm \mu^\mp$ events with $\mu^\pm \mu^\pm e^\mp$ events. We now study such a discrimination quantitatively based on the previous simulation results.

The essence of this statistical analysis is to perform a hypothesis test on two competing hypotheses, namely, the nature of sterile neutrino can be either Dirac or Majorana.
We follow here the frequentist approach and consider two benchmark scenarios:  $m_N^{} = 20~\text{GeV}$ and $50~\text{GeV}$. The observables are the numbers of events in the various final states $\ell^\pm_{} \ell^\pm_{} \ell^{\prime \mp}_{}$, which are treated as the data set. In Tables \ref{tb:mN20} and \ref{tb:mN50}, we have listed the numbers of \emph{expected} events for the case of the mixing angles $|U_{eN}|^2=|U_{\mu N}|^2 = 1\times 10^{-6}$ and $U_{\tau N}^{} = 0$. Apparently, one can easily obtain the expected numbers of events for other mixing angles by simple rescaling. Moreover, by further applying statistical fluctuations, one can generate realistic pseudo-data (simulated data) samples. For convenience, in addition to the \emph{disparity factor}  $r = |U_{Ne}/U_{N\mu}|^2$, we introduce another parameter, $s$, a \emph{normalization factor} defined as: 
\begin{equation}
s \equiv  2\cdot10^6\, \times  \frac{|U_{Ne}^{}U_{N\mu}^{}|^2}{ |U_{Ne}|^2+ |U_{N\mu}|^2 },
\label{normalization_factor}
\end{equation}
 i.e. $s$ is a measure of the smallest of the two mixings. Thus, for the case given in Tables \ref{tb:mN20} and \ref{tb:mN50}, we have $s = r = 1$.

Having obtained the realistic pseudo-data samples,
we next fit them with the two competing hypotheses. A $\chi^2_{}$ function is built so as to characterize how well the pseudo-data sets are described within a given hypothesis,
\begin{eqnarray}
\chi^2_{H} = -2 \underset{s,r \subset H}{\text{min}} \left \{ \text{ln} \left( \prod_i ~ \text{Poiss} \left [ N_i^{\text{expc}}(s,r;H), N_i^{\text{obs}} \right ] \right) \right \},
\end{eqnarray}
where $H$ stands for the Dirac or Majorana hypothesis of sterile neutrinos, $i$ denotes a particular trilepton final state, and $\text{Poiss}(N^{\text{expc}}_{}, N^{\text{obs}}_{})$ is the probability of observing $N^{\text{obs}}_{}$ events in Poisson statistics when the number of expected events is $N^{\text{expc}}_{}$. The above definition also involves a minimization procedure that is taken upon the free parameters i.e., $s$ and $r$ in the hypothesis.

\begin{figure}[h]
\centering
\includegraphics[scale=0.7]{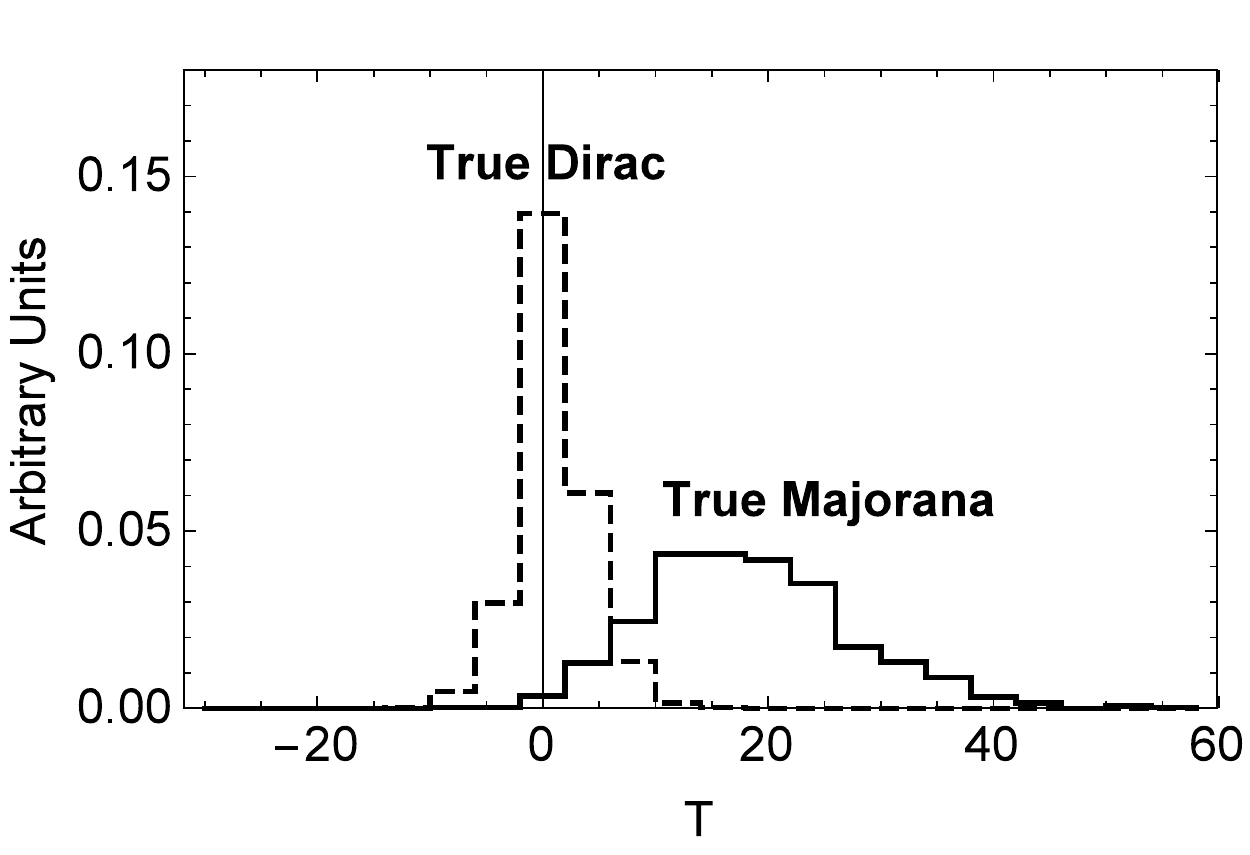}
\caption{Distributions of test statistic $T$ for the case with $s = 1$ and $ r= 5$, given the true nature of sterile neutrinos is Dirac (dashed) or Majorana (solid).}
\label{fg:test_statistic}
\end{figure}

To quantify which hypothesis is more favorable, we then define a test statistic  $T$ as
\begin{eqnarray}
T = \chi^2_{\text{Dirac}} - \chi^2_{\text{Maj}}.
\end{eqnarray}
Because of statistical fluctuations, the obtained values of $T$ for different pseudo-data samples, which correspond to the same expected numbers of events (or the same set of input parameters $s$ and $r$), can be different. In Figure \ref{fg:test_statistic} we show the probability distributions of the test statistic $T$ for 1000 pseudo-data samples that all have $s = 1$ and $r = 5$, assuming either the true Dirac (dashed) or Majorana (solid) nature of sterile neutrinos. As expected, when sterile neutrinos are truly Dirac particles, we obtain almost equally well fitting for both hypotheses so that $T$ is centered around 0. Namely, in this case no discrimination power can be obtained. However, when the true nature of sterile neutrinos is Majorana, the Majorana hypothesis has a better fit (a smaller $\chi^2_{}$ value), resulting in a positive value of $T$. Statistical fluctuation causes the spread of two distributions, and the level of their overlap determines the confidence level of discriminating these two hypotheses.

\begin{figure}[h]
\centering
\includegraphics[scale=0.65]{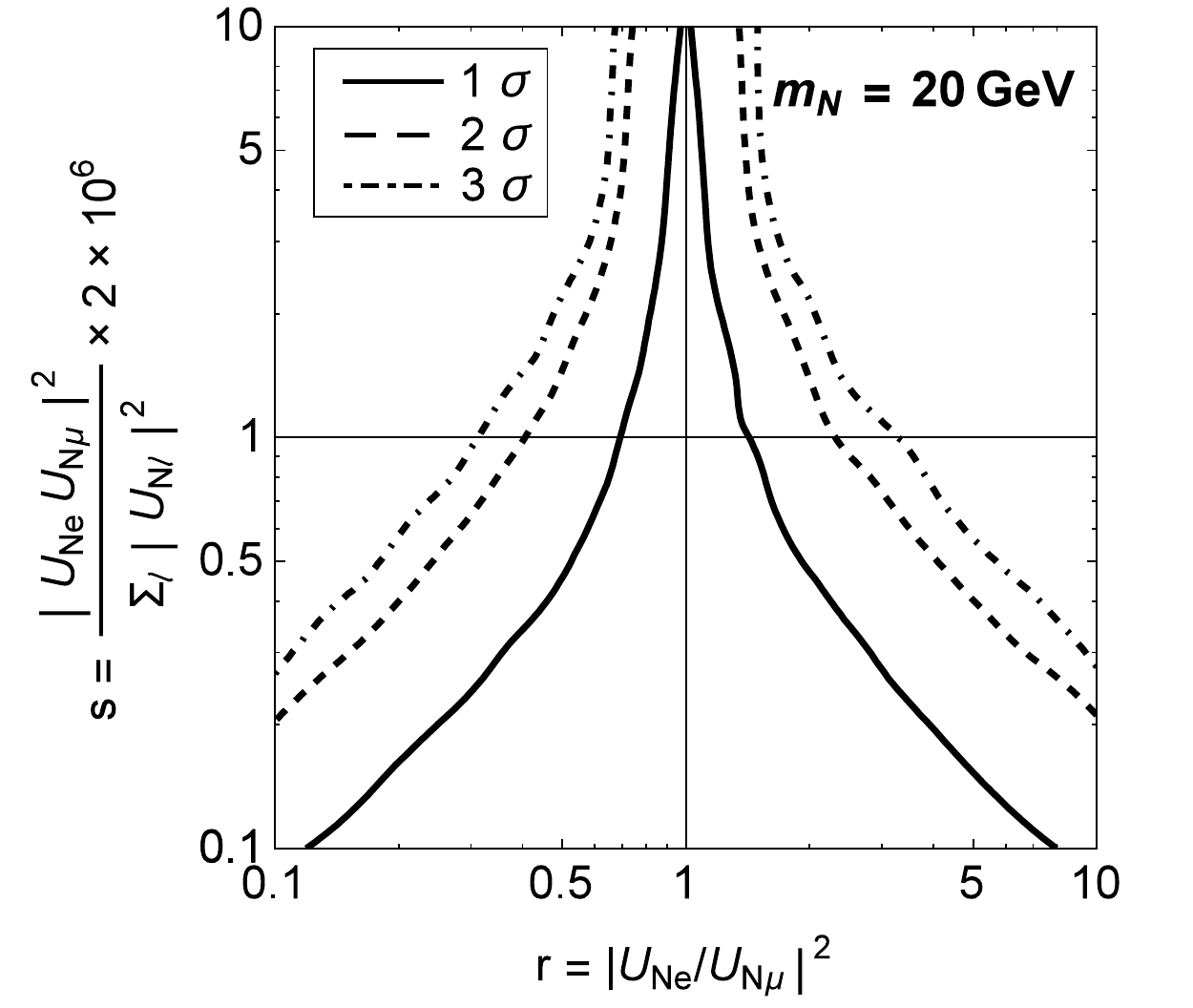}~ \includegraphics[scale=0.65]{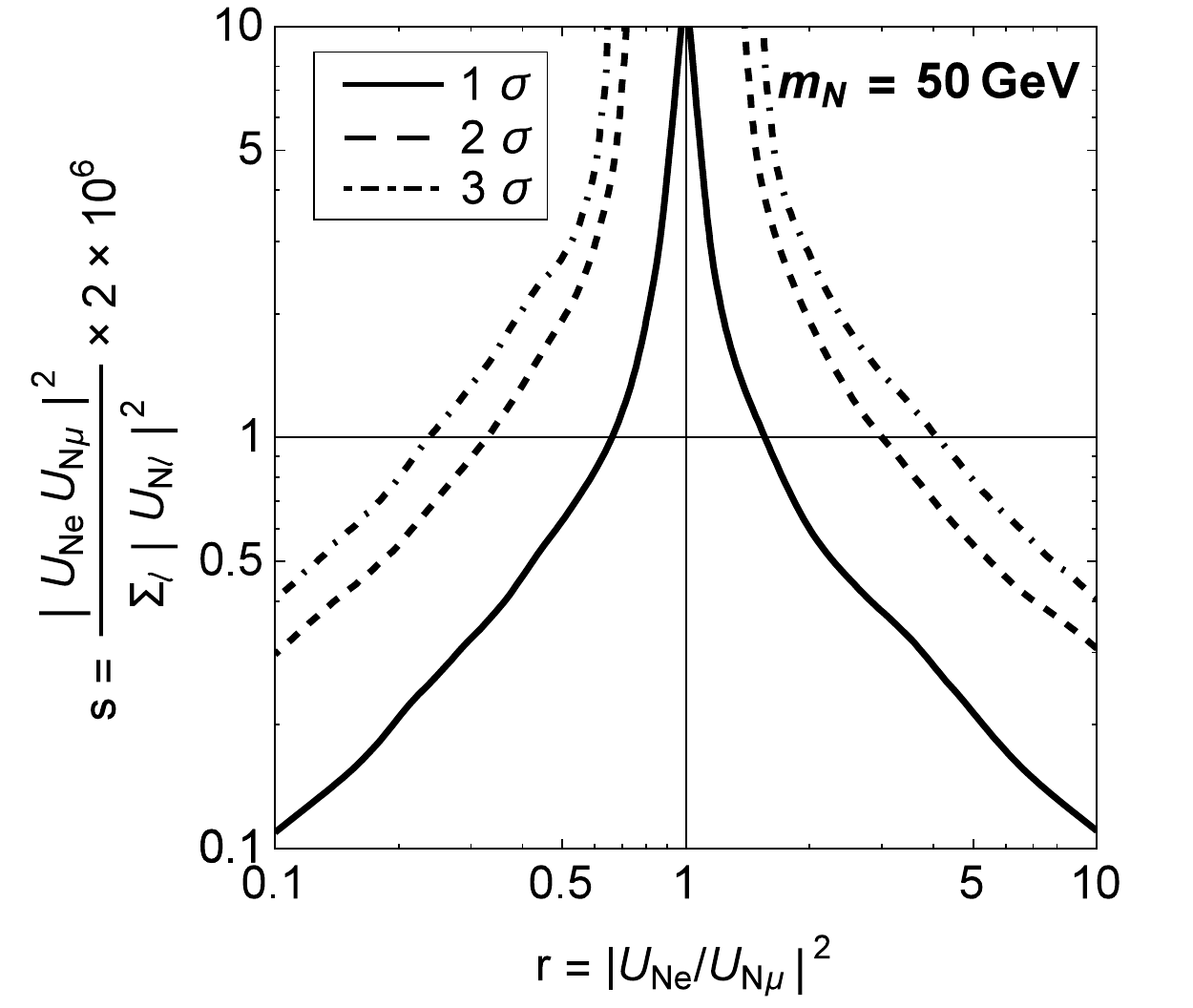}
\caption{Confidence levels of excluding Dirac type given the true nature of sterile neutrinos is of Majorana type, for two benchmark scenarios of $m_N^{} = 20~\text{GeV}$ (left) and $50~\text{GeV}$ (right). Horizontal and vertical axes respectively denote the true values of the mixing angle ratio $r$ and the normalization factor $s$ in logarithmic scales. Solid, dashed and dot-dashed curves correspond to the $1\sigma$, $2\sigma$ and $3\sigma$ confidence levels of excluding the Dirac type, respectively. 
}
\label{fg:scenario_1}
\end{figure}

To simplify the complexities caused by the statistical fluctuations, we consider a ``median" discrimination. Namely, for the true Dirac case, where the distribution of $T$ is sharply peaked at zero, we therefore choose $T = 0$ as the median possible value of $T$.
Then, given the true Majorana nature of sterile neutrinos, the confidence level of excluding the Dirac hypothesis can be quantified as $1-\alpha$, where $\alpha $ is the probability of explaining the true Majorana nature with the wrong Dirac one. In terms of Figure \ref{fg:test_statistic}, this $\alpha$ is the area under the blue curve for $T < 0$.

Finally, in Figure \ref{fg:scenario_1}, we present our main point:  the numerical results on the discrimination of Majorana vs. Dirac neutrinos based on the disparity of the mixings $|U_{Ne}|^2$ and $|U_{N\mu}|^2$. We do this for two benchmark scenarios $m_N^{} = 20~\text{GeV}$ (left) and $50~\text{GeV}$ (right). The main question is to determine how far from unity the disparity factor $r$ has to be in order to tell a Majorana character apart from Dirac. Horizontal and vertical axes respectively denote the true values of the disparity factor $r$ and the normalization factor $s$, which are used in generating the pseudo-data samples. Solid, dashed and dot-dashed curves correspond to excluding the Dirac hypothesis given the true Majorana nature at $1\sigma$, $2\sigma$ and $3\sigma$ levels, respectively. As one can see, for both benchmark scenarios, at least a $3\sigma$ level exclusion can be reached for disparities as mild as e.g. $r \lesssim 0.7$ (or $1/r \lesssim 0.7$), provided $s \gtrsim 5$. For smaller $s$ (smaller mixings), which means fewer events, one clearly requires larger values of $r$ to reach the same level of discrimination; in the same way, as $r$ approaches 1, larger values of $s$ are required as it becomes more and more difficult to exclude the Dirac case. 
Further discriminating power will require additional information from the spectral distributions of the produced leptons, an issue that we will discuss in a later work.

\section{Conclusions}
\label{sec:summary}

In this work we focus on the question of determining the nature of sterile neutrino with mass below $M_W^{}$ at the LHC. Because of such a low mass for the neutrino, the conventional same sign dilepton plus jet search for Majorana sterile neutrinos at the LHC suffers from the issue of insufficient phase space for final state leptons and jets passing the necessary detector cuts. Therefore, we choose to study the alternative tri-lepton search channel, and to reduce SM backgrounds we further require no opposite-sign same-flavor lepton pairs in the final state. Although this tri-lepton search is ideal for such a low mass sterile neutrino search, it turns out to be a non-trivial task of pinning down the underlying nature of sterile neutrinos, as the final state neutrinos or anti-neutrinos, which carry valuable information about lepton number, are not detected at current colliders.

A simple scenario is identified in this tri-lepton search, so that a discrimination on the nature of sterile neutrinos can be possible. This fortunate scenario could arise if the underlying nature of sterile neutrinos are of Majorana type, and their mixing angles with electrons and muons are different enough. We find that in this fortunate scenario, one is able to exclude the Dirac sterile neutrino case by simply counting and then comparing the numbers of events in the $e^+_{} e^+_{} \mu^-_{}$ and $\mu^+_{} \mu^+_{} e^-_{}$ channels (or the corresponding charge-conjugated ones). We perform a careful collider simulation for this scenario. According to our statistical analysis, at the $14~\text{TeV}$ LHC with an integrated luminosity of $3000~\text{fb}^{-1}$, at least a $3\sigma$ level of exclusion on the Dirac case can be achieved, depending on the size and disparity of the two relevant mixing parameters.  For example, such is the case for $s \gtrsim 5$ and $r \lesssim 0.7$ (or $1/r \lesssim 0.7$), where $s = 2\cdot 10^6  \times
|U_{Ne}^{}U_{N\mu}^{}|^2/ \left(
|U_{N e}|^2 +|U_{N\mu}|^2
\right) $, and $r = |U_{Ne}/U_{N\mu}|^2$. For other values of mixings, see Fig.~\ref{fg:scenario_1}.
Therefore, in the current collider search for sterile neutrinos with masses below $M_W^{}$, a quick check of this scenario via the above method can be very rewarding, as if sterile neutrinos were indeed in this mass range and their  mixing to electron and muon flavor were different enough, we might know the nature of those sterile neutrinos quite shortly.

\section*{Acknowledgements}

J.Z. thanks Prof. Shun Zhou for helpful discussions, encouragement and support, Brian Shuve for sharing details on fake lepton simulation, and Yan-dong Liu for help on using $\texttt{DELPHES 3}$; K.W. thanks Prof. Cai-Dian L$\rm \ddot{u}$ for his help; and C.D. thanks Juan C. Helo for helpful discussions. 
J.Z. was supported in part by the Innovation Program of the Institute of High Energy Physics under Grant No. Y4515570U1; K.W. by the International Postdoctoral Exchange Fellowship Program (No.90 Document of OCPC, 2015);  C.S.K. by the NRF grant funded by the Korean government of the MEST (No. 2011-0017430) and (No. 2011-0020333); and C.D. by Chile grants Fondecyt No.~1130617, Conicyt FB0821 and ACT1406. 


\appendix
\section{Validation for Fake Lepton Simulation}
\label{sec:FL}

In this appendix, we intend to present our validation results for the fake lepton simulation used in this work. We follow closely the same validation done in Ref.~\cite{Izaguirre:2015pga}, and find out that using their modelling parameters the simulation results can indeed be consistent with the experimental results given in Ref.~\cite{Chatrchyan:2014aea}. Specifically, we take $r_{10}^{} = 1$, $\mu = 0.5$, $\sigma = 0.3$ and $\epsilon_{200}^{} = 4.6 \times 10^{-3}_{}$. In fact, the suggested mistag rate of $\epsilon_{200}^{} = 4.6 \times 10^{-3}_{}$ coincides with the ``rule-of-thumb" introduced in Ref.~\cite{Sullivan:2010jk}, i.e., isolated electrons and muons from heavy-flavor decay are about $1/200$ times the rates of $b$ and $c$ quark production. For the other input parameters of $r_{10}^{}$, $\mu$ and $\sigma$, the authors of Ref.\cite{Izaguirre:2015pga} find that varying them does not substantially change the fitting to the data, provided the overall fake efficiency of $\epsilon_{200}^{}$ remains fixed.

\begin{figure}[h]
\centering
\includegraphics[scale=0.6]{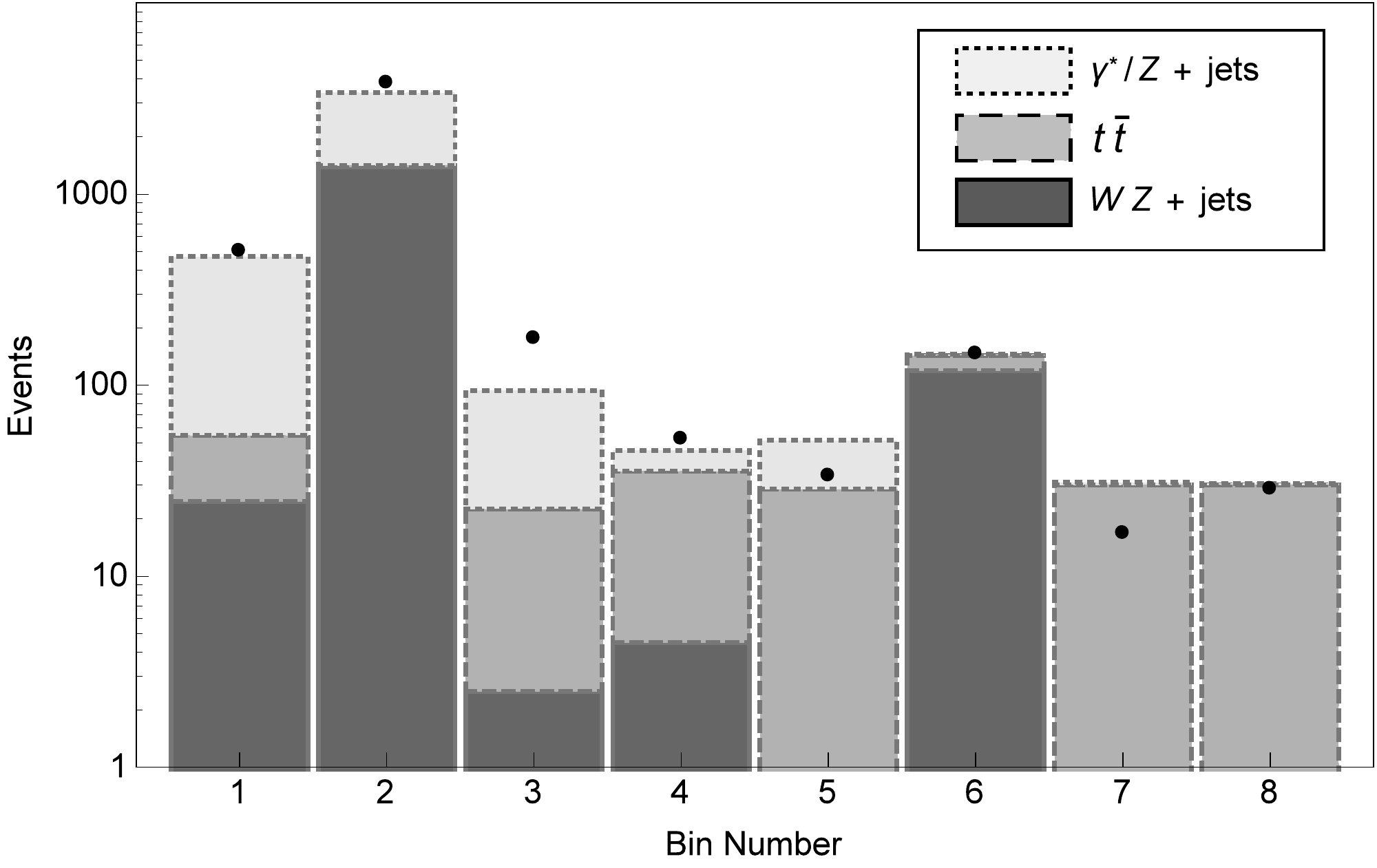}
\caption{Validation results for fake lepton simulation. Black dots indicate experimental results in Ref. \cite{Chatrchyan:2014aea}. Our simulated results for $\gamma^*_{}/Z$+jets, $t\bar{t}$ and $WZ$+jets are given by up light gray bars, middle brown bars and bottom pink bars, respectively. Eight bin categories are: (1) 0-bjet, 1-OSSF, $M_{\ell^+_{}, \ell^-_{}}^{} < 75 ~{\rm GeV}$; (2) 0-bjet, 1-OSSF, $|M_{\ell^+_{}, \ell^-_{}}^{} - M_Z^{} | < 15 ~{\rm GeV}$; (3) 0-bjet, 1-OSSF, $M_{\ell^+_{}, \ell^-_{}}^{} >105 ~{\rm GeV}$; (4) 0-bjet, 0-OSSF; (5-8) are the same as the first four bins, but with at least one b-jet.}
\label{fg:validation}
\end{figure}

Our validation results are shown in Figure \ref{fg:validation}. Each bin represents an event category according to the CMS trilepton search given in Ref.~\cite{Izaguirre:2015pga}, namely, (1) 0-bjet, 1-OSSF, $M_{\ell^+_{}, \ell^-_{}}^{} < 75 ~{\rm GeV}$; (2) 0-bjet, 1-OSSF, $|M_{\ell^+_{}, \ell^-_{}}^{} - M_Z^{} | < 15 ~{\rm GeV}$; (3) 0-bjet, 1-OSSF, $M_{\ell^+_{}, \ell^-_{}}^{} >105 ~{\rm GeV}$; (4) 0-bjet, 0-OSSF; (5-8) are the same as the first four bins, but with at least one b-jet. The actual experiment results are indicated by black dots, while our simulated results are given by up light gray bars, middle brown bars and bottom pink bars for processes of $\gamma^*_{}/Z$+jets, $t\bar{t}$ and $WZ$+jets, respectively. As one can see, our results agree with the experimental results reasonably well within the statistical uncertainty, especially in Bin-4, whose selection criteria mostly resemble the ones in our main text. Moreover, a good agreement with the results given in Figure 10 of Ref.~\cite{Izaguirre:2015pga} is also found, although in some bins we differ in the individual fractions of events from different processes.

\section{Determining $m_N^{}$ and Identifying the Origins of Same Sign Dileptons}
\label{sec:SL_order}

We here introduce a method of identifying the origin of the same sign dileptons in the final state, namely, whether they come from the prompt decay of the W-boson or from the sterile neutrino $N$. Since this method requires the knowledge of the sterile neutrino mass $m_N^{}$, we therefore also discuss possible ways of determining $m_N^{}$ as we proceed. In fact, $m_N^{}$ may be known as a by-product of our procedure.

Our method starts with the full reconstruction of the four-momentum of the missed neutrino in the final state. This can be approximately achieved by assuming that the s-channel produced W-boson has a small transverse momentum, and requiring the invariant mass of all final state particles equals the W-boson mass. Because of the quadratic nature of the invariant mass equation, two possible solutions of the longitudinal momentum of neutrino will be found.

With these two possible solutions for the four-momentum of the neutrino, we can then reconstruct \emph{four} possible values of $m_N^{}$ for a given event, as there exists another two-fold ambiguity in choosing one lepton from the same-sign dilepton. Among these four possible values we know one of them must be quite close to the true value of $m_N^{}$, while no such a connection exists for the other three reconstructed values of $m_N^{}$. Thus, if we plot the distribution of such reconstructed $m_N^{}$'s for all events, we should expect to observe a peak at the true value of $m_N^{}$.

\begin{figure}[h]
\centering
\includegraphics[scale=0.65]{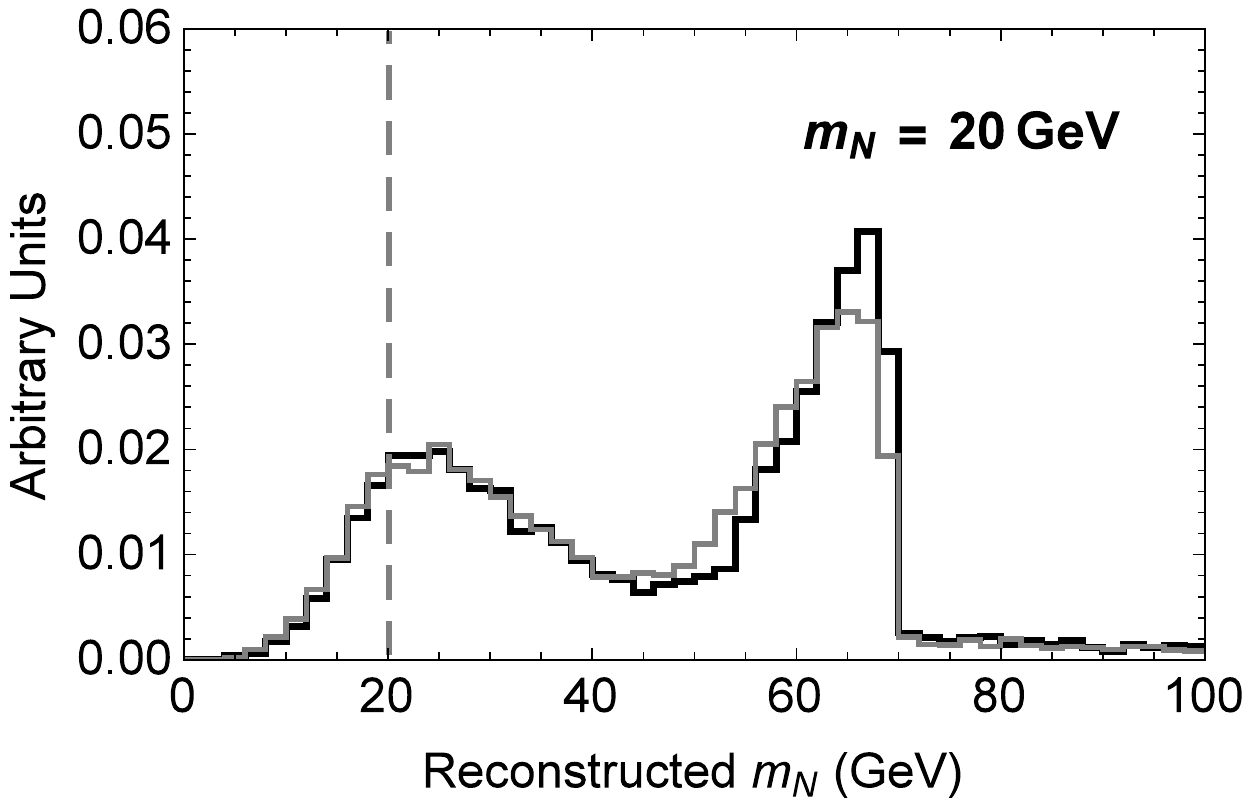}
\includegraphics[scale=0.65]{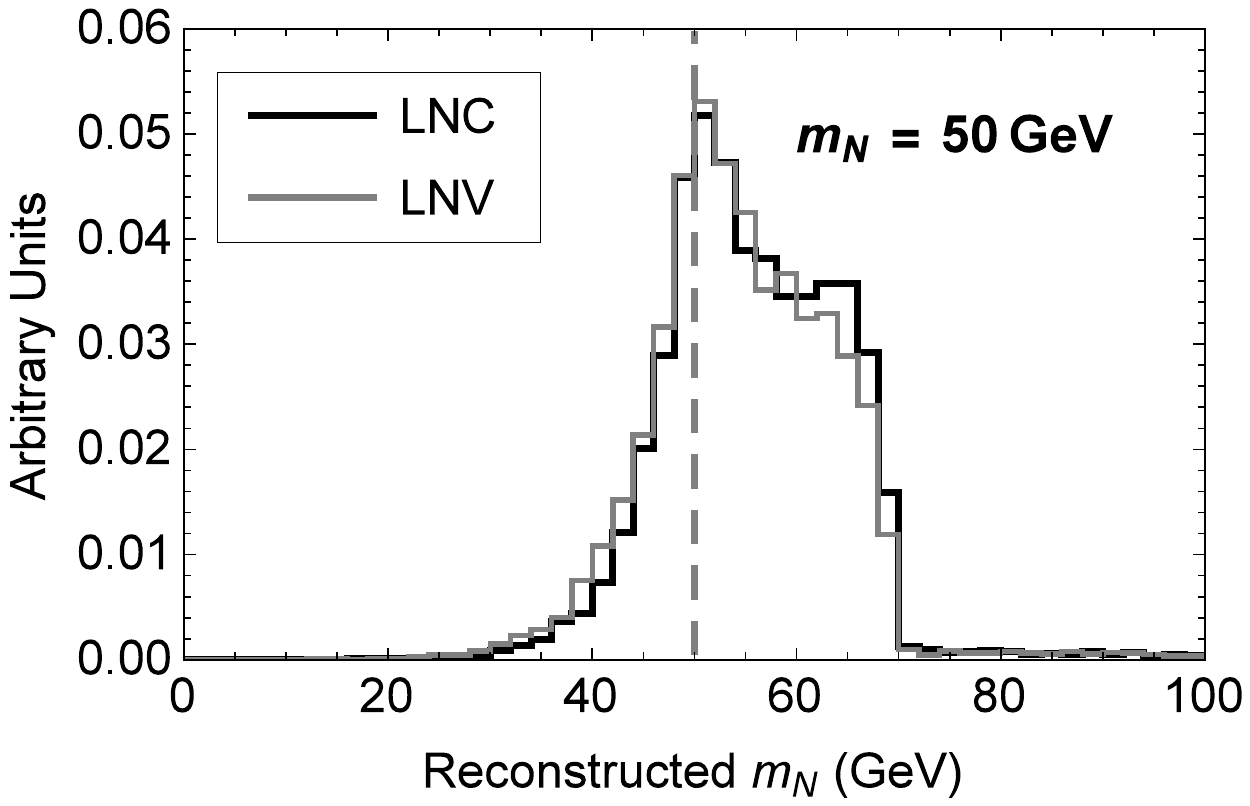}
\caption{Distributions of the reconstructed $m_N^{}$ using the method given in Appendix \ref{sec:SL_order} for the signal process in the $e^+_{} e^+_{} \mu^-_{}$ final state with $m_N^{} = 20~\mathrm{GeV}$ (left) and $m_N^{} = 50~\mathrm{GeV}$ (right). The black and gray solid curves represent the LNC and LNV sub-processes, respectively.}
\label{fg:mN}
\end{figure}

In Figure \ref{fg:mN} we provide the distributions of the reconstructed $m_N^{}$ for the signal process in the $e^+_{} e^+_{} \mu^-_{}$ final state with $m_N^{} = 20~{\rm GeV}$ (left) and $m_N^{} = 50~{\rm GeV}$ (right), and the LNC and LNV sub-processes are distinguished by black and gray solid curves, respectively. As one can see, in the $m_N^{} = 50~{\rm GeV}$ case one indeed observes a peak at the true value of $m_N^{}$. However, for the $m_N^{} = 20~{\rm GeV}$ case two peaks are found, and the correct peak is the one that is less sharp.\footnote{The presence of a sharp peak at $\sim 65~{\rm GeV}$ is due to the fact that with $m_N^{} = 20~{\rm GeV}$, the lepton from the prompt W-boson decay would have an energy close to $60~\text{GeV}$. Then, if this lepton were used in the reconstruction of $m_N^{}$, a peak near $\sim 65~{\rm GeV}$ could be observed.} Tentatively, one may then conclude that such a reconstruction method is not applicable to the lower value $m_N^{}$ case. However, one can actually exclude the possibility of the sharper peak being the correct one by employing the argument that if the true value of $m_N^{}$ were at the sharp peak, $\sim 65~{\rm GeV}$, we then should observe a single peak instead of two, according to our example of $m_N^{} = 50 ~{\rm GeV}$. Therefore, even for the lower value $m_N^{}$ case one can still extract some useful information about $m_N^{}$ via such a reconstruction, although not as precise as the large $m_N^{}$ case.

A few more comments on determining $m_N^{}$ are in order. First, just within the above reconstruction method, one can actually obtain a more precise value of $m_N^{}$ by performing a fit to the distribution reconstructed from the real data. Second, even if the above reconstruction method is still not good enough for the low $m_N^{}$ case, one can improve the determination of $m_N^{}$ by combining it with other methods. For example, in this low $m_N^{}$ case one actually knows with a great certain that the lepton with a smaller $p_T^{}$ in the same sign dileptons comes from the decay of $N$. Thus, by plotting the invariant mass of this lepton and the opposite sign lepton one should observe an end point at the true value of $m_N^{}$. The authors in Ref.~\cite{Izaguirre:2015pga} indeed adopt this method of identifying the origins of the same sign dileptons. Such a method, however, is only applicable to the low $m_N^{}$ case. When $m_N^{} \sim m_W^{}$, it is hard to know the origin of the same sign dileptons by comparing their transverse momenta. In contrast, our reconstruction method actually performs better in this large value $m_N^{}$ case. In this sense, these two approaches are complementary. Lastly, we here ignore the impact of background on the reconstruction method, since the distribution of the reconstructed $m_N^{}$ for the background processes has a fixed shape, which therefore can be removed from the distribution of the real data in the first place.

Having discussed possible ways of determining $m_N^{}$, we assume that $m_N^{}$ is known from now on. The correct origin of the same sign dileptons is then identified as follows. In each event, among four possible reconstructed values of $m_N^{}$, the one that is closest to the true value of $m_N^{}$ is assumed to be correct one. Thus, leptons in that combination are taken as leptons from the decay of the sterile neutrino $N$. Meanwhile, by doing this we can also know the four-momentum of the final state neutrino. In other words, all four-momenta of the final state particles are now known. This fact is crucial when discussing the discrimination between Dirac and Majorana sterile neutrinos by building distributions of various kinematic variables.

Finally, it should be noted that because of  detector effects, the off-shell production of the $W$-boson and its possibly non-negligible transverse momentum due to initial state radiation, just using this method, it is not possible to reconstruct the correct origin of the same-sign dileptons at $100\%$ confidence level. For the $m_N^{} = 20 ~{\rm GeV}$ case our correctness is about $80\%$, similar to the $p_T^{}$ method used in Ref.~\cite{Izaguirre:2015pga}, while for the $m_N^{} = 50 ~{\rm GeV}$ case a correctness around $75\%$ is obtained, in contrast to $\sim 65\%$ using the $p_T^{}$ method.


\end{document}